\documentclass[12pt]{article}
\usepackage{amssymb,amsmath,comment,mathtools}
\usepackage[dvipdfmx]{graphicx}
\usepackage{subfigmat}
\usepackage{braket}
\usepackage{hyperref} 	  		
\usepackage{epsfig}
\usepackage{here}
\usepackage{color}
\usepackage{bm}
\graphicspath{{./Figures/}}

\newcommand{\ba}{\begin{align}}
\newcommand{\ea}{\end{align}}

\def\nn{\nonumber}

\def\bea{\begin{eqnarray}}
\def\eea{\end{eqnarray}}
 \makeatletter
\def\alt{\mathrel{\mathpalette\gl@align<}}
\def\agt{\mathrel{\mathpalette\gl@align>}}
\def\gl@align#1#2{\lower.6ex\vbox{\baselineskip\z@skip\lineskip\z@
\ialign{$\m@th#1\hfil##\hfil$\crcr#2\crcr\sim\crcr}}} \makeatother
\renewcommand{\thefootnote}{\fnsymbol{footnote}}
\begin{document}
\begin{flushright}
\end{flushright}
\vspace*{1.0cm}

\begin{center}
\baselineskip 20pt 
{\Large\bf 
Higgs Portal Majorana Fermionic Dark Matter with the Freeze-in Mechanism
}
\vspace{1cm}

{\large 
Junpei Ikemoto${}^{a,b}$, Naoyuki Haba${}^{b,c}$, Shimizu Yasuhiro${}^{b,c}$ and Toshifumi Yamada${}^{d}$
} \vspace{.5cm}

{\baselineskip 20pt \it
${}^{a}$Institute of Science and Engineering, Shimane University, Matsue 690-8504, Japan\\
${}^{b}$Department of Physics, Osaka Metropolitan University, Osaka 558-8585, Japan \\
${}^{c}$Nambu Yoichiro Institute of Theoretical and Experimental Physics (NITEP),
Osaka Metropolitan University, Osaka 558-8585, Japan\\
${}^{d}$Department of Physics, Yokohama National University, Yokohama 240-8501, Japan
}

\vspace{.5cm}

\vspace{1.5cm} {\bf Abstract} \end{center}

We consider a minimal model of fermionic dark matter, in which the Majorana fermion dark matter (DM) $\chi$ couples with the Standard Model (SM) Higgs field $H$
 through a higher-dimensional term $-{\cal L}\supset H^\dagger H \bar{\chi}\chi/\Lambda$, where  $\Lambda$  is the cutoff scale.
We assume that $\Lambda$ is sufficiently large that DM particles are not in thermal equilibrium with the SM Particles throughout the history of the Universe.
Hence, DM particles are produced only by the freeze-in mechanism.
Through a numerical analysis of the freeze-in mechanism, 
we show contour plots of the DM relic abundance for various values of the DM mass, reheating temperature and the cutoff scale.
We  obtain an upper bound of the DM mass and cutoff scale from contour plots on ($m_\chi, \Lambda$)-plane. We also consider the direct DM detection for the parameter regions where the DM relic abundance is consistent with the experimental values. We find that the spin-independent cross section for the elastic scattering with a nucleon is below the current experimental upper bound.

\thispagestyle{empty}

\newpage
\renewcommand{\thefootnote}{\arabic{footnote}}
\setcounter{footnote}{0}
\baselineskip 18pt
\section{Introduction}

The SM of particle physics can accurately explain most of the current experiments. However, the SM cannot explain several observations and experimental facts,
such as the existence of DM particles in the present universe.
The existence of DM has been firmly supported by astrophysical and cosmological observations.
Assuming the validity of General Relativity,
DM is observed to be ubiquitous in gravitationally collapsed structures of sizes ranging 
from the smallest known galaxies \cite{smallest galaxies} to galaxies of size comparable to the Milky Way \cite{Milky Way},
to groups and clusters of galaxies \cite{clusters of galaxies}.
In such collapsed structures,
 the existence of DM is inferred directly using tracers of mass enclosed within a certain radius, such as stellar velocity dispersion,
  rotation curves in axisymmetric systems,
 the virial theorem, gravitational lensing,
  and measures of the amount of non-dark,
   i.e., baryonic, mass such as stellar number counts and tracers of gas density such as X-ray emission \cite{inferred direct }.
The relic abundance of DM  in the present universe is determined \cite{pdg Astrophysic}:
\bea\label{eq:observedValue}
	\Omega_{\rm DM}h^2 = 0.12\pm0.0012
	\quad.
\eea 
The generation of DM particles in the early universe can proceed via thermal or non-thermal production, or both.
The former is sometimes referred to as the {\it freeze-out} mechanism, in which some large amount of DM exists in the early universe and decreases to the currently observed amount due to thermal equilibrium with the SM particles.
The latter is sometimes referred to as the {\it freeze-in} mechanism,  in which   the non-thermal-equilibrium particles (here, DM particles) are produced 
 from thermal equilibrium with the SM particles \cite{sample3}.

Although we know nothing about the properties of DM particles, such as their mass, spin and couplings to the SM particles,
we would like to know the interaction between the DM and the SM particles.
We consider the Higgs-portal Majorana fermion DM scenario, in which the Majorana fermion DM $(\chi)$ and the SM Higgs boson $(H)$ interact through a higher-dimensional operator,  
 $-{\cal L}\supset H^\dagger H \bar{\chi}\chi/\Lambda$,
 where $\Lambda$ is a cutoff scale.
If $\Lambda$ is sufficiently large, the DM particle is not in thermal equilibrium with the SM particles throughout the history of the Universe.
Hence, the DM particles are produced only by the freeze-in mechanism.
  Ref. \cite{sample} considered the freeze-in mechanism in the regions of
  $\Lambda=M_{\rm planck}$ and $m_\chi > 10^{10}$ GeV based on the WINMPzilla hypothesis and calculated  the DM relic abundance.
  In this paper, we analyze the DM relic abundance over a wider range of $\Lambda$  and  lighter DM masses. 
  In addition, we examine the direct DM detection rate for parameter regions where the DM relic abundance is consistent with the experimental values.
Note that when we evaluate the DM relic abundance,
it is necessary to set the initial conditions for the temperature at which DM production begins, 
  because this term is not renormalizable at high temperature as the early universe $\Lambda \ll T$. 
Therefore,
 we define the initial condition for dimensionless quantity $x=m/T_{\rm RH}$,
 where $T_{\rm RH}$ is the reheating temperature.
 We consider $T_{\rm RH}$ to be larger than the electroweak phase transition temperature of $10^2$ GeV, which guarantees baryogenesis that produces enough baryons to produce the number of baryons observed in the present universe. 
 We numerically calculate the relic abundance of the Majorana fermionic DM by the freeze-in mechanism and show contour plots of the DM relic abundance for various the DM mass, reheating temperature and the cutoff scale.
 We  obtain the upper bounds of the DM mass and cutoff scale from contour plots on ($m_\chi, \Lambda$)-plane.
 We calculate the direct DM detection for the parameter regions where the DM relic abundance is consistent with the experimental values and find that the spin-independent cross section for the elastic scattering with a nucleon is below the current experimental upper bound.

This paper is organized as follows. 
In Section 2,
 we set up the Higgs portal Majorana fermionic dark matter. 
In Section 3,
 we explain the freeze-in mechanism with Higgs portal DM in our model.
Direct dark matter detection with the DM-nucleus elastic scattering cross section is discussed in Section 4.
We conclude in Section 5.


\section{Model}

We introduce a gauge-singlet Majorana fermion, $\chi$, to the SM field content.
We also introduce a $Z_2$ symmetry, under which $\chi$ is odd and the SM fields are even. The relevant part of the Lagrangian involving $\chi$ reads
\bea
{\cal L} \ \supset \ \frac{i}{2}\bar{\chi}\gamma^\mu\partial_\mu\chi - \frac{m}{2} \bar{\chi}\chi - \frac{\bar{\chi}\chi H^\dagger H}{\Lambda} 
	\quad ,
\eea
where $H$ denotes the SM Higgs field, and $\Lambda$ is a cutoff scale of the model.
Because of the $Z_2$ parity, the neutral particle $\chi$ with mass $m$ becomes stable and provides a natural DM candidate.\footnote{
We hereafter denote the DM particle also by $\chi$.
}

Since $\bar{\chi}\chi H^\dagger H$ is the lowest dimensional operator that couples the DM with SM particles,
 we concentrate on the DM production through this operator in the radiation-dominated Universe.
As shown in the next section,
 we restrict $\Lambda$ to the case of $\Lambda \gg T_{\rm RH}$, where $T_{\rm RH}$ is a reheating temperature and starting temperature of the freeze-in mechanism,
 so that the amplitudes do no diverge. 


\section{Dark Matter Production}

We consider a radiation-dominated era of the Universe, where the SM particles are in thermal equilibrium at temperature $T$.
We define $x=m/T$, and write the yield of the $\chi$ at the temperature $T$ as $Y(x)$,
 and the yield of the DM if it were in thermal equilibrium as $Y_{\rm eq}(x)$.
The Boltzmann equation that regulates $Y(x)$ is given by
\bea \label{eq:boltzmann eq}
	\frac{{\rm d}Y(x)}{{\rm d}x} \ = \ -\frac{\langle\sigma v\rangle}{x^2}\frac{S(m)}{H(m)}\left(Y(x)^2-Y_{\rm eq}(x)^2\right)
	\quad,
\eea 
 where $\langle\sigma v\rangle$ is the thermal average of the cross section times velocity of the process $\chi \chi \leftrightarrow \ $(SM particle pair),
 $S(T)$ is the entropy density at temperature $T$,
  and $H(T)$ is the Hubble rate at $T$.
The thermal average $\Braket{\sigma v}$ of the cross section $ \chi \chi \leftrightarrow HH$ is given by the following integral expression\cite{boltzman}:
\bea
	\Braket{ \sigma v }
		= \frac{g_\chi^2}{64 \pi^4} \left( \frac{m_\chi}{x}  \right) \frac{1}{ n_{\rm eq}^2 } 
				\int^\infty_{4m^2_\chi} ds ~s \sqrt{s-4m^2_\chi}  (\sigma v) K_1\left( \frac{x\sqrt s}{m_\chi}  \right)
	\quad ,
\eea
where $g_\chi$=2 counts the physical degree of freedom of the Majorana fermion $\chi$,
 $n_{\rm eq} = S(m_\chi) Y_{\rm eq} / x^3$,
 $(\sigma v)=W^{ {\rm 2-body } }_{ {\rm ij } } / s$, and $K_1$ is the modified Bessel function of the second kind of order 1.
Since the Higgs portal amplitude is given by $ |\mathcal M_{\chi \chi \leftrightarrow HH}|^2 = 2s/\Lambda^2$,
\bea
	s(\sigma v)
		= \frac{ \sqrt  s \sqrt{s-4m_\chi^2} }{32\pi \Lambda^2}
	\quad .
\eea
And then, 
 the thermal average $\Braket{\sigma v}$  in this model can be written as
\bea
	\Braket{ \sigma v }
		=\frac{ 
			\pi
			\left[
				G^{3,0}_{1,3}
				\left( x^2\left|  
				\begin{array}{cc}
   					1 \\
   					0,2,3
				\end{array}
				\right.
				\right)
				-2x^4 K_2(2x)
			\right]}{
			32  \Lambda^2 x^4 K_2(x)^2 
			}
	\quad ,
\eea
where we use $n_{eq} = m_\chi^3 g_\chi K_2(x) / ( 2\pi^2 x)$ .
Here,  
$G^{m,n}_{p,q}
				\left( z\left|  
				\begin{array}{cc}
   					a_1, \cdots , a_p\\
   					b_1, \cdots , b_q
				\end{array}
				\right.
				\right)
$
is the Meijer G-function, defined by a line integral in the complex plane
\bea
G^{m,n}_{p,q}
	\left( z\left|  
	\begin{array}{cc}
   		a_1, \cdots , a_p\\
   		b_1, \cdots , b_q
	\end{array}
	\right.
	\right)
		=
	\frac{1}{2\pi i}
	\int_\gamma
	\frac{
	\prod^m_{j=1} \Gamma(b_j - s)\prod^n_{j=1} \Gamma(1 - a_j + s)
	}{
	\prod^p_{j=n+1} \Gamma(a_j - s)\prod^q_{j=m+1} \Gamma(1 - b_j + s)
	}	
	x^s ds
	\quad ,
\eea
where $\Gamma(z)$ is the gamma function and $\gamma$ indicates the appropriate contour \cite{meijer g-function}.

\begin{figure}[t]
  \begin{subfigmatrix}{2}
    \subfigure[]{\includegraphics[scale=0.8]{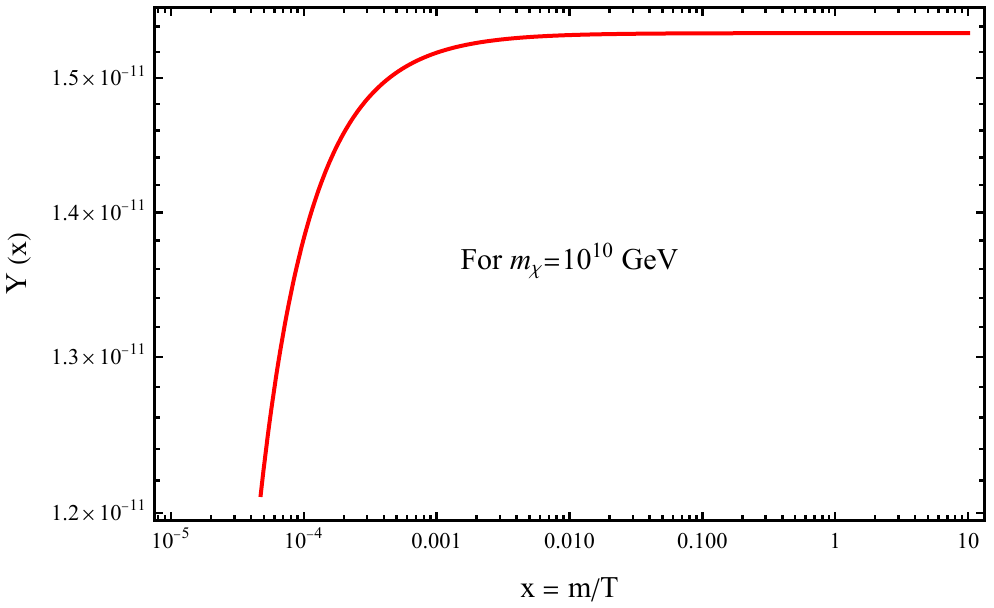} }
    \subfigure[]{\includegraphics[scale=0.8]{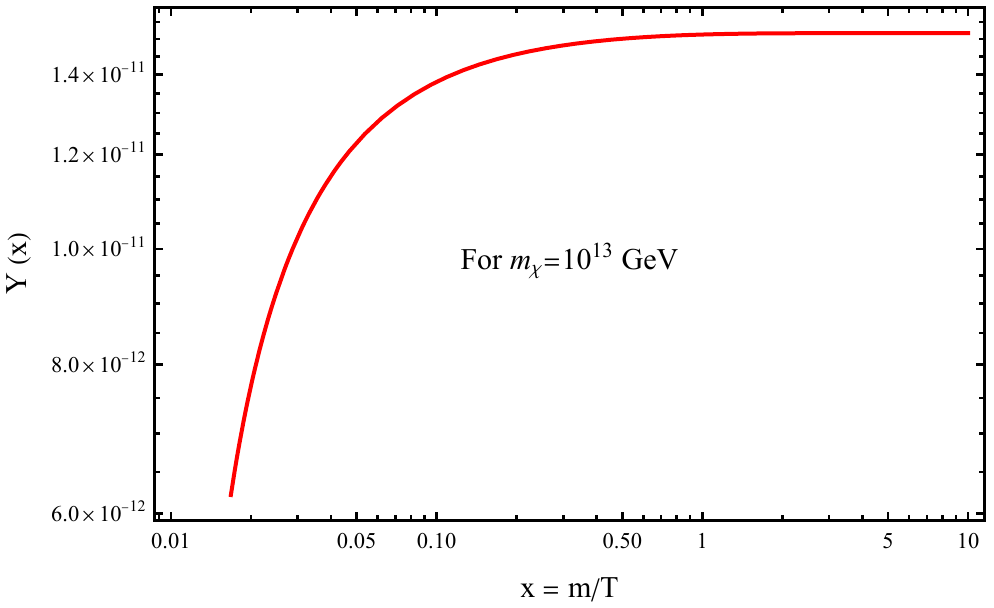} }
  \end{subfigmatrix}  
  \caption{
     The evolution of the yield of $\chi$ as a function of $x=m/T$.
  In solving the Boltzmann equation, we fixed $\Lambda = 10^{20}$ GeV, $T_{\rm RH}=10^{15}$ and $m_\chi= 10^{10}$ GeV in (a) and $m_\chi=10^{13}$ GeV in (b). 
  }
\label{fig:approBoltzmannSample}
\end{figure}

By numerically solving the Boltzmann equations Eq.(\ref{eq:boltzmann eq}) with the initial condition $Y(x_{\rm RH}=m/T_{\rm RH}) = 0$ for various values of $m_\chi$, $\Lambda$ and $T_{RH}$, 
we evaluate the yield of $\chi$ at the present universe $ Y(x \to \infty) $.
In our analysis, we assume $ \Lambda \gg T_{\rm RH} $ to ensure that $\chi$ is not in thermal equilibrium with the SM particle.
We comment that we only consider a DM mass above half the Higgs boson mass. Then the decay of the Higgs boson into two DM's is kinematically forbidden and it is not necessary to include the process $h \to \chi \chi$ ($h$ denotes the Higgs boson) in the analysis of DM relic abundance.
Fig.\ref{fig:approBoltzmannSample} shows the evolution of the yield of $\chi$ as a function of $x$.

The relic abundance of the $\chi$ at the present universe is given by
\bea 
	\Omega_{\chi} h^2 = \frac{m_\chi S_0 Y(\infty)}{\rho_c / h^2} \quad ,
\eea 
where $S_0=2890~{\rm cm}^{-3}$ is the entropy density at the present universe and 
$\rho_c / h^2=1.05 \times 10^{-5}~{\rm GeV /  cm}^3 $ is the quantity of the critical density divided by the scaling factor for the Hubble expansion rate.
In the case of Fig.\ref{fig:approBoltzmannSample}(a) with our model,
 $\Omega_{\chi} h^2 \sim 4 \times 10^7$.
This value does not match the relic abundance of the DM at the present universe, Eq.(\ref{eq:observedValue}).
Then we consider a contour plot to find parameters that match the observation value.
\begin{figure}[H]
  \begin{subfigmatrix}{2}
    \subfigure[]{\includegraphics[scale=1]{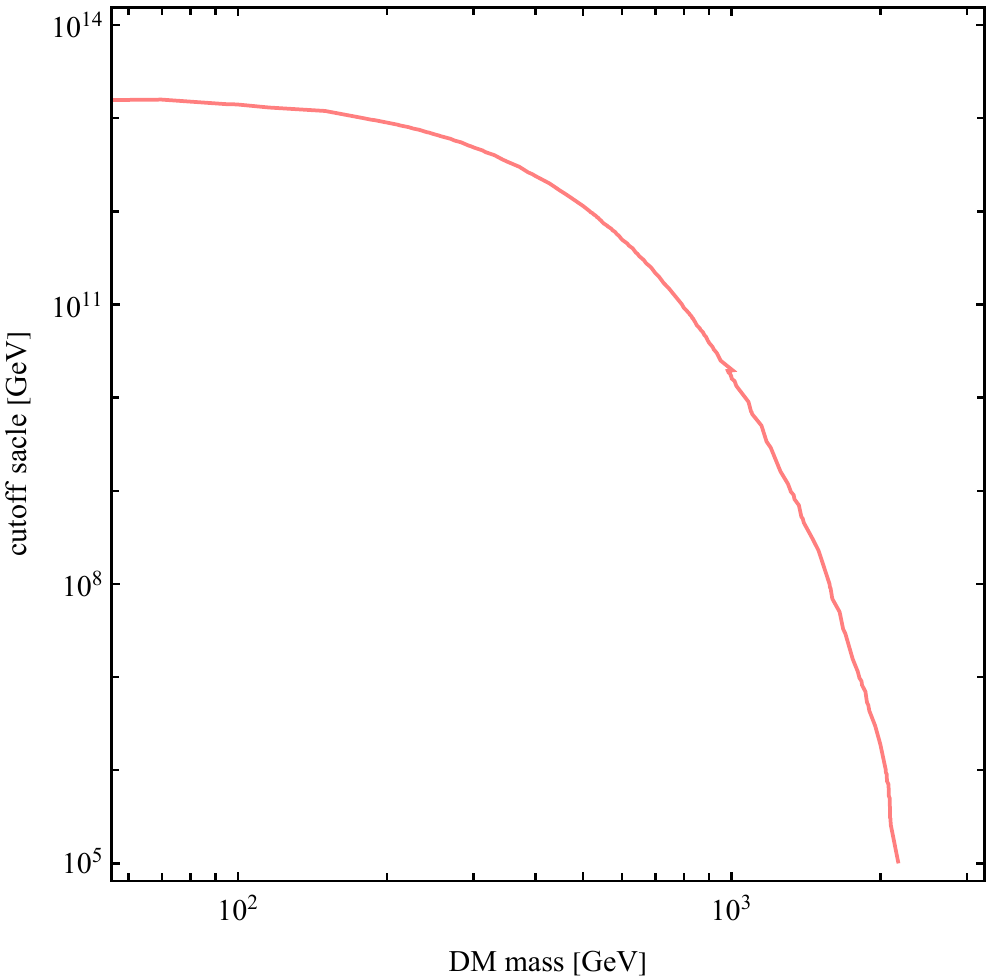} }
    \subfigure[]{\includegraphics[scale=1]{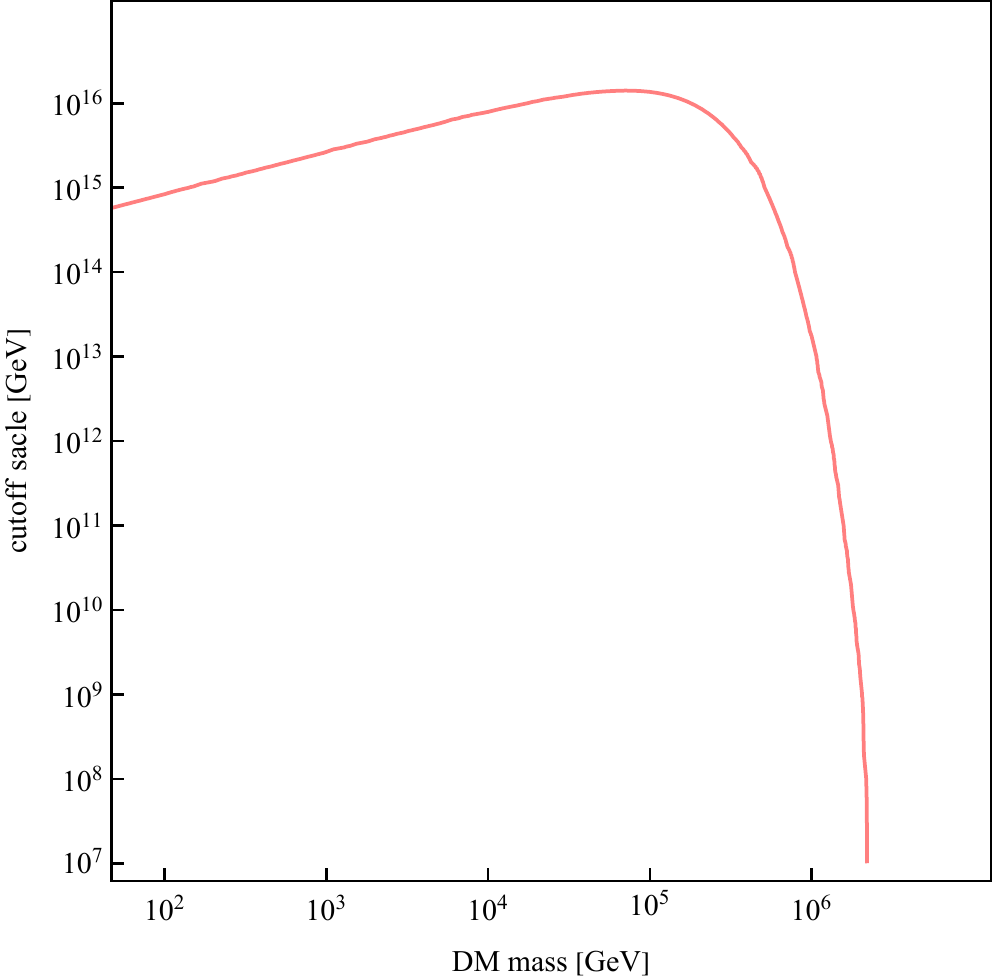} }
    \subfigure[]{\includegraphics[scale=1]{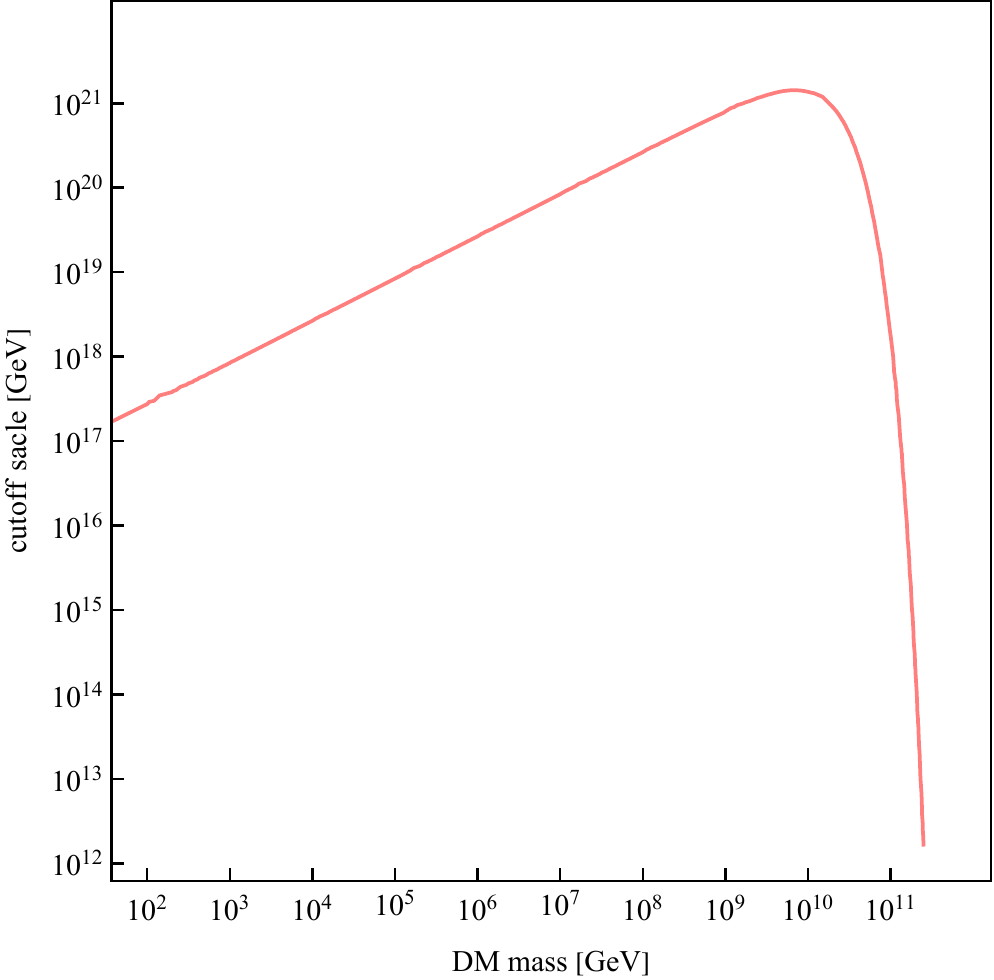} }
    \subfigure[]{\includegraphics[scale=1]{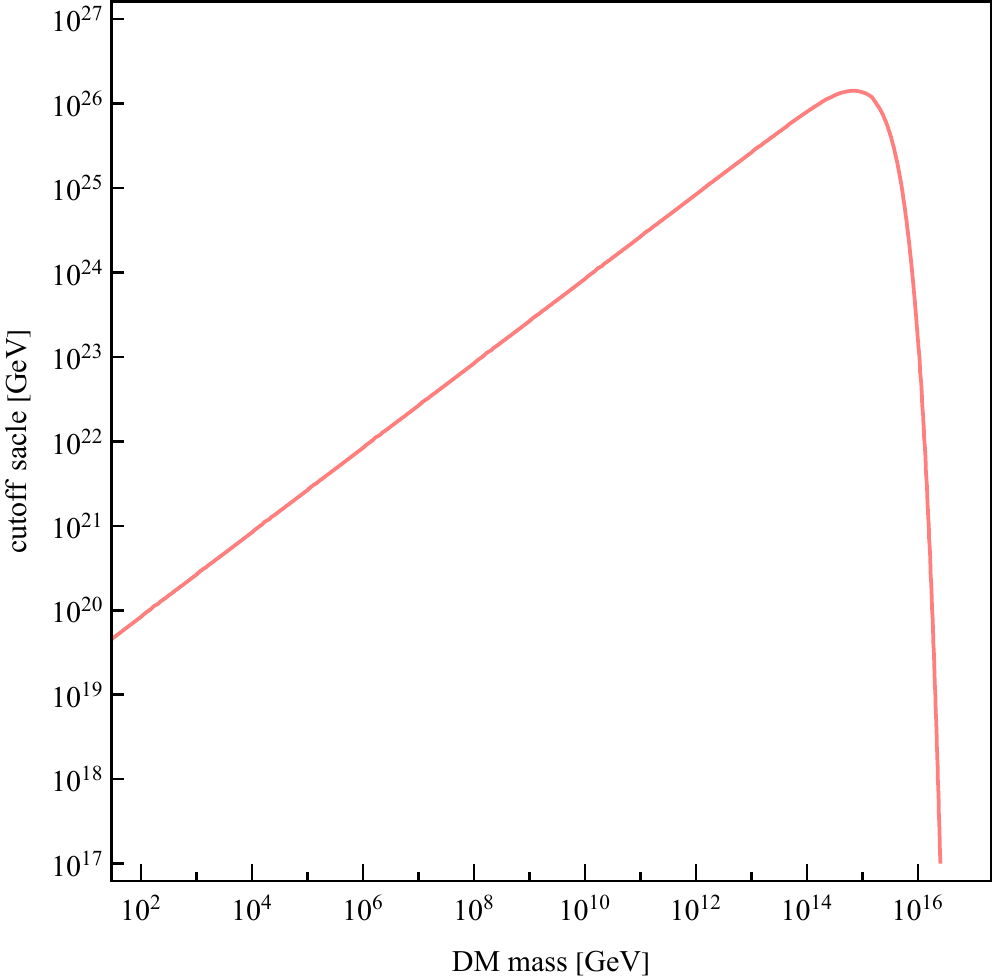} }
  \end{subfigmatrix}  
  \caption{
  The contour plots for fixed $T_{\rm RH}$ that match the observed value $\Omega_{\rm DM} h^2 = 0.12$. 
  The Fig.\ref{fig:Contourplot} (a), (b), (c) and (d) are contour plots fixed at $T_{RH}=10^2, 10^5, 10^{10}$ and $10^{15}$ GeV, respectively.
}
  \label{fig:Contourplot}
\end{figure}
Fig.\ref{fig:Contourplot} shows contour plots for $\Omega_\chi h^2=0.12$ on the $(\Lambda , m_\chi)$-plane with fixed $T_{RH}$.
In this analysis, 
 Fig.\ref{fig:Contourplot} includes above the Planck scale ($10^{19}$ GeV),
 but the regions may not have physical meaning. 
From Fig.\ref{fig:Contourplot}, we can see that the contour plots fall rapidly to a certain $m_\chi$.
This is because the smaller the momentum of the SM Higgs in the thermal equilibrium than  $m_\chi$,
the more difficult it is for the SM Higgs to produce $\chi$.
In order to produce the $\chi$,
therefore,
the freeze-in mechanism in this model requires a smaller cutoff scale.   
We can determine upper bounds of $m_\chi$ and $\Lambda$ from figures on the $(\Lambda , m_\chi)$-plane with fixed $T_{RH}$.
The upper bound of $m_\chi$ apears around $10$ times $T_{RH}$.
This is because the $Y(x)$ is suppressed by the Boltzmann factor $\exp ( {-E_\chi/T} )$
 when $x\sim 10$.
In the case of $T_{RH}=100$ GeV,
 we can see that the upper bound of $m_\chi$ and $\Lambda$ is 2100 GeV and approximately $10^{13}$ GeV,
 respectively.  

\begin{figure}[t]
  \begin{subfigmatrix}{2}
    \subfigure[]{\includegraphics[scale=1]{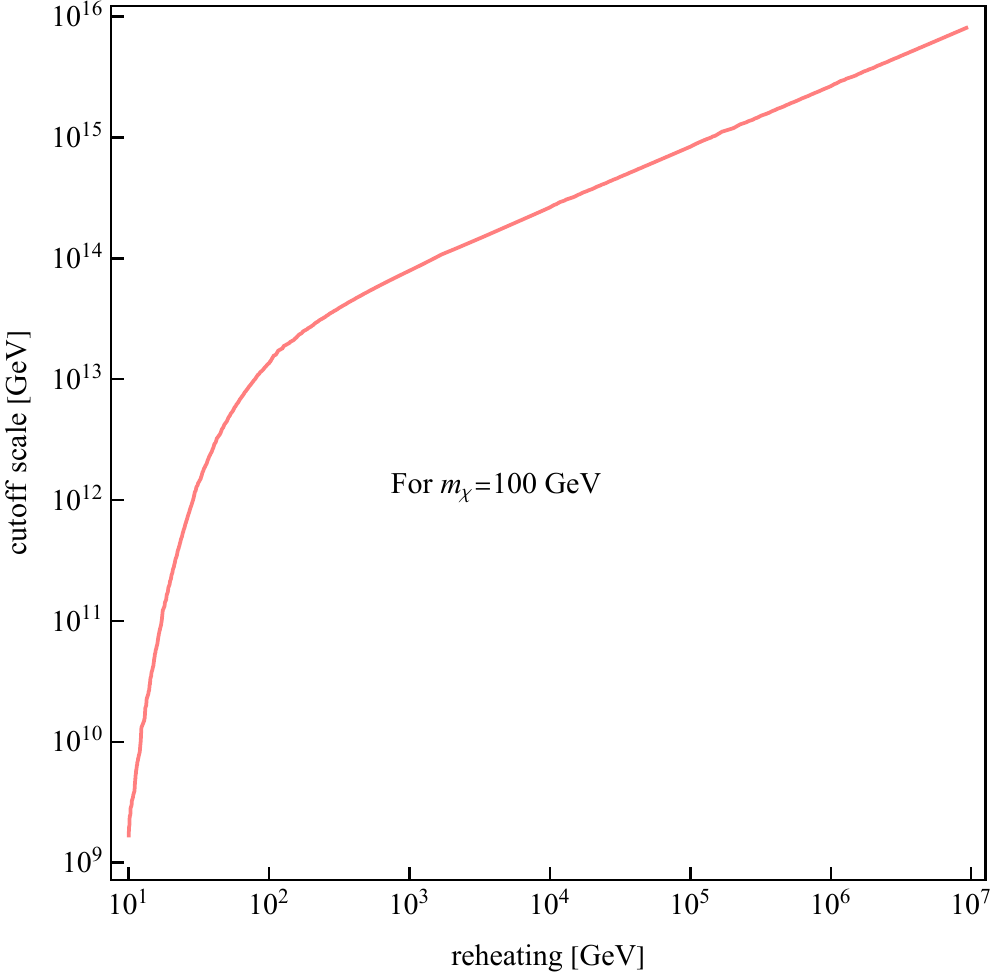} }
    \subfigure[]{\includegraphics[scale=1]{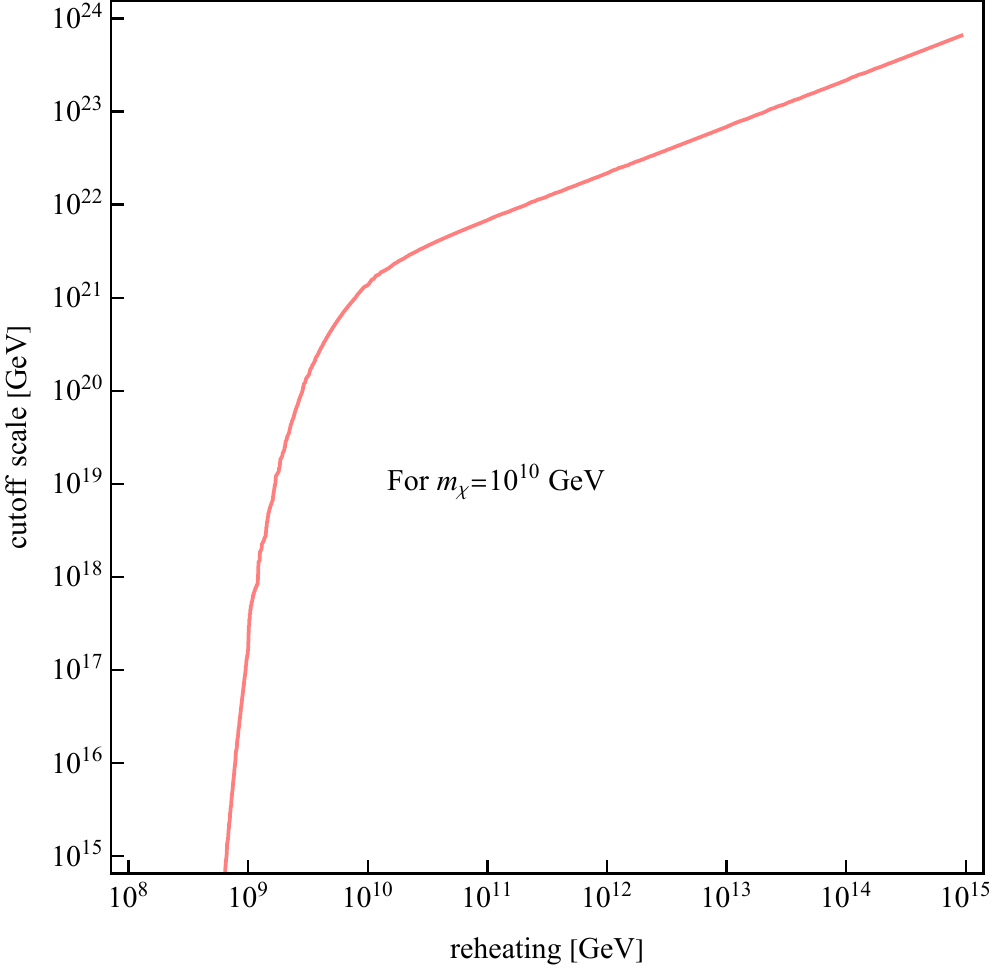} }
  \end{subfigmatrix}  
  \caption{
  The contour plots for fixed $m_\chi$ that match the relic abundance of $\chi$ at the present universe. 
  The Fig.\ref{fig:Contourplot2} (a) and (b) are contour plots fixed at  $m_\chi=10^2$ and $10^{10}$ GeV, respectively.} 
  \label{fig:Contourplot2}
\end{figure}

Similarly,
Fig.\ref{fig:Contourplot2} shows that the contour plots damping to a certain $\Lambda$ in order to produce sufficient $\chi$. 

\section{Direct Dark Matter Detection}
DM $\chi$ couples to the SM quarks through the SM Higgs with a cutoff scale $\Lambda$. 
We calculate only spin-independent effective couplings between $\chi$ and SM quarks,
because $\bar \chi \chi H^\dagger H $ is CP invariant.
The SM Higgs exchange interaction reads
\bea
	\mathcal L^h_{\rm{eff} }
		=m_Q ( \bar Q Q )( \bar \chi \chi ) 
			\left( \frac{-1}{16\pi^2} \right) 
			\frac{1}{m_h^2}\frac v \Lambda
	\quad .	 	
\eea  
Let us calculate the cross section for DM-nucleon elastic scattering. 
We first estimate the following matrix elements:
\bea
	\Braket{N| m_Q \bar Q Q |N}	 	
		\equiv m_N f^N_{TQ}
		\qquad
		(Q = u,d,s,c,b,t)
		\quad ,
\eea
where $N$ represents a proton, $p$, 
or a neutron, $n$. 
A lattice calculation \cite{lattice} reports that
\bea
	f^p_{Tu} + f^p_{Td} 
		=f^n_{Tu} + f^n_{Td}
		\backsimeq 0.056,
		\qquad
		|f^N_{Ts}|<0.08 
	\quad .
\eea
We hereafter adopt these values, but we set $f^N_{Ts} = 0 $ to make our analysis conservative.
Operators involving heavy quarks, $c, b, t$ give rise to an effective coupling to gluons through a triangle diagram \cite{triangle diagram}, 
and in the heavy quark limit we have
\bea
	\Braket{N| m_Q \bar Q Q |N}
		\simeq 
			\Braket{N| 
				\left ( -\frac{1}{12\pi} \alpha_s \right)
				G_{\mu\nu}G^{\mu\nu}
			|N}
	\qquad {\rm for} ~ Q=c,b,t
	\quad .
\eea
Using the trace of the QCD energy-momentum tensor given by
\bea 
	\theta^\mu_\mu
		= \sum_{Q=u,d,s,c,b,t} m_Q \bar Q Q - \frac{7\alpha_s}{8\pi}G_{\mu\nu}G^{\mu\nu}
		\simeq \sum_{Q=u.d.s} m_Q \bar Q Q - \frac{9\alpha_s}{8\pi}G_{\mu\nu}G^{\mu\nu}
	\quad ,  
\eea
we obtain 
\bea
	m_N \simeq 
		\Braket{N| \sum_{Q=u,d,s} m_Q \bar Q Q |N} + \Braket{N| \left( -\frac{9\alpha_s}{8\pi} G_{\mu\nu}G^{\mu\nu} \right) |N}
	\quad .
\eea
We thus find for $ Q=c,b,t $,
\bea
	\Braket{N | m_Q \bar Q Q | N}
		&\simeq& \Braket{N | \left( -\frac{2\alpha_s}{24\pi}G_{\mu\nu}G^{\mu\nu} \right) | N}  \nn \\
		&\simeq& \frac{2}{27} (1 - f^N_{Tu} - f^N_{Td} - f^N_{Ts} ) m_N
	\quad .
\eea
Using the above expressions, the spin-independent cross section for the elastic scattering of $\chi$ with a nucleon is given by
\bea\label{eq:DM-necleon scattering}
	\sigma^{\rm SI}
		&=& \frac 1 \pi \frac{ m^2_N m^2_{\chi}  }{ ( m_N+m_{\chi} )^2 } m^2_N
			\left\{
				f^N_{Tu} + f^N_{Td} + f^N_{Ts} + \frac 2 9(1 - f^N_{Tu} - f^N_{Td} - f^N_{Ts})
			\right\}^2 
		\nn \\
		& \times&
			 \left( \frac{1}{16\pi^2}\frac{1}{m_h^2}\frac v\Lambda  \right)^2
		\quad ,
\eea
where $m_N \simeq 1 $GeV is the mass of a nucleon (proton or neutron).

\begin{figure}[t]
	\centering
	{\includegraphics[scale=1]{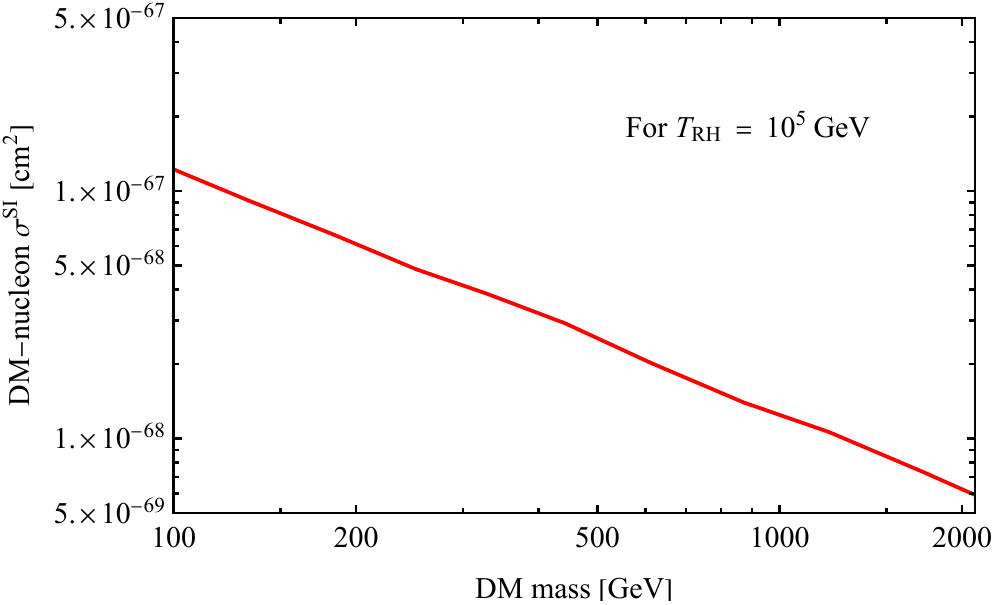} }
  \caption{
    The theoretical prediction of the DM-nucleon $\sigma^{\rm SI}$ in the parameter regions where the DM relic abundance is consistent with the experimental value for $T_{\rm RH}=10^5$ GeV (Fig.\ref{fig:Contourplot}(b)).}
    
  \label{fig:DM direct ver2}
\end{figure}

Liquid xenon time projection chambers  are the leading technology in the search for a large variety of dark matter particle candidates.
Now is the time to design the ultimate next-generation DM experiment in order to probe the widest possible range of DM candidates
\cite{next generation direct detection exprement }.
To evaluate whether the DM-nucleon $\sigma^{\rm SI}$ will be observed at future experiments,
 we substitute $m_\chi$ and $\Lambda$ that agree with the observed value Eq.(\ref{eq:observedValue}) in Fig.\ref{fig:Contourplot} into Eq.(\ref{eq:DM-necleon scattering}). 
Then, we compare the theoretical predictions of this model with the expected upper limit of the next-generation experiment.  
Observing the DM-nucleon $\sigma^{\rm SI}$ in a future experiment will provide information  not only the DM mass but also the reheating temperature.
\begin{figure}[t]
  \centering
  {\includegraphics[scale=1]{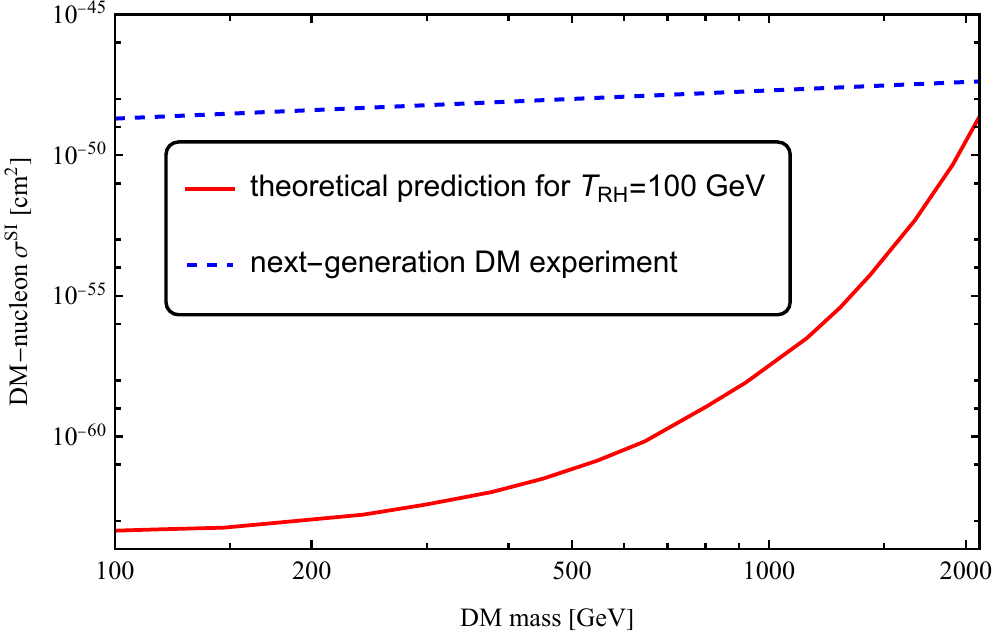} }
  \caption{
  The comparison of the DM-nucleon $ \sigma^{\rm SI}$ between the theoretical prediction of our model,
  shown in solid red line,
  and the expected upper limit for exposures of 1000 ton-year of the next-generation experiment,
  shown by dashed blue line.
 }
  \label{fig:direct detection}
\end{figure}
Fig.\ref{fig:direct detection} shows the DM-nucleon $ \sigma^{\rm SI}$ as a function of $m_\chi$ for $T_{RH}=10^2$ GeV, and the cutoff $\Lambda$ is chosen as the DM relic abundance is consistent with the experimental value ( Fig.\ref{fig:Contourplot}(a)~).
The DM-nucleon $ \sigma^{\rm SI}$ is found to be slightly lower than the sensitivity of the next-generation experiment, reaching a maximum at $m_\chi=2100$ GeV, which is the upper mass bound for fixed $T_{\rm RH}=100$ GeV.
Fig.\ref{fig:DM direct ver2} shows the DM-nucleon $ \sigma^{\rm SI}$ for $T_{\rm RH}=10^5$ GeV, which is much smaller than that for $T_{\rm RH}=100$ GeV. This is because the larger cutoff scale is allowed for the larger values of $T_{\rm RH}$ so that  the $ \sigma^{\rm SI}$ is suppressed by the cutoff scale. Thus, in our model, the $ \sigma^{\rm SI}$ is found to be  maximum when
  $T_{\rm RH}=100$ GeV and $m_\chi=2100$ GeV.

Conversely, 
the above results suggest that the lower $T_{RH}$ becomes, the lower cut-off scale $\Lambda$ , and as a result, the $\sigma^{\rm SI}$ becomes large enough to allow future observations.
In fact,
 in the regions below $T_{\rm RH} \ll 100$ GeV,
 the $\sigma^{\rm SI}$ for $T_{\rm RH}=\mathcal O (10 {~\rm MeV})$ and $m_\chi= \mathcal O (1 ~{\rm MeV})$ is expected to be observed in the future.
\cite{lower Limits Reheating, lower Limits Reheating direct detedtion}

\section{Conclusion}
We have considered a minimal model of fermionic dark matter, in which the Majorana fermion dark matter (DM) $\chi$ couples with the Standard Model (SM) Higgs field $H$  through a higher-dimensional term $-{\cal L}\supset H^\dagger H \bar{\chi}\chi/\Lambda$, where  $\Lambda$  is the cutoff scale.
We assume the cutoff scale $\Lambda \gg T_{RH}$ in order to avoid a divergence arising from the nonrenormalizable interaction $H^\dagger H \bar\chi  \chi /\Lambda$ at $T \gg \Lambda$
and ensure that DM was not in thermal equilibrium during DM production.
Therefore, in our model,
 DM-genesis occurred not by the freeze-out mechanism but by the freeze-in mechanism.

 We numerically solved the Boltzmann equations with the initial condition $Y(x_{\rm RH}=m/T_{\rm RH})=0$ for various values of  $m_\chi$ 
 , $\Lambda$ and $T_{\rm RH}$. We examined the parameter regions of $m_\chi$ and $\Lambda$ where  the DM relic abundance is consistent with  the observed value $\Omega_{\rm DM}h^2 = 0.12$, and determined the upper bound of $m_\chi$ and $\Lambda$.
 These upper bounds are caused from the difficulty that the smaller momentum of the SM Higgs in the thermal equilibrium than $m_\chi$  rarely produces $\chi$.
Thus,
 the cutoff energy $\Lambda$ must rapidly decrease in order to sufficiently produce an abundance of $\chi$.

We also considered the direct DM detection for the parameter regions where the DM relic abundance is consistent with the experimental values. We showed that the spin-independent cross section $\sigma^{\rm SI}$ for the elastic scattering with a nucleon is below the current experimental upper bound and there is no experimental constraint at present. 
  Since the larger cutoff scale is allowed for the larger values of $T_{\rm RH}$ by the constraint on the DM relic abundance, the $ \sigma^{\rm SI}$ is suppressed by the cutoff scale. Therefore, we found  that the $ \sigma^{\rm SI}$ becomes  maximum for $T_{\rm RH}=100$ GeV, $m_\chi=2100$ GeV.

\section*{Acknowledgments}
This work is partially supported by Scientific Grants by the Ministry of Education,
Culture, Sports, Science and Technology of Japan, No. 21H00076 (NH) and No. 19K147101 (TY).


\end{document}